\documentclass[a4paper,fleqn,usenatbib,useAMS]{mn2e}

\usepackage{graphicx}	
\usepackage{amsmath}	
\usepackage{amssymb}	
\usepackage{multicol}        
\usepackage{bm}		
\usepackage{pdflscape}	

\usepackage[english]{babel}
\usepackage{blindtext}

\usepackage[T1]{fontenc}
\usepackage{ae,aecompl}

\usepackage{newtxtext,newtxmath}

\title[CO first overtone and mass accretion rate in MYSOs]{Blinded by the light: on the relationship between CO first overtone emission and mass accretion rate in massive young stellar objects}
\author[J.~D.~Ilee et al.]{%
J.~D.~Ilee$^{1}$\thanks{Contact e-mail: \href{mailto:jdilee@ast.cam.ac.uk}{jdilee@ast.cam.ac.uk}},
R.~D.~Oudmaijer$^{2}$,
H.~E.~Wheelwright$^{3}$ and
R.~Pomohaci$^{2}$
\vspace{0.4em}
\\
$^{1}$Institute of Astronomy, Madingley Road, Cambridge CB3 0HA, UK\\
$^{2}$School of Physics and Astronomy, EC Stoner Building, University of Leeds, Leeds, LS2 9JT, UK\\
$^{3}$Max-Planck-Institut f\"{u}r Radioastronomie, Auf dem H\"{u}gel 69, 53121, Bonn, Germany\\
}

\date{Accepted 2018 April 3. Received 2018 April 3; in original form 2017 October 27}

\pubyear{2018}

\begin{document}
\label{firstpage}
\pagerange{\pageref{firstpage}--\pageref{lastpage}}
\maketitle

\begin{abstract}

To date, there is no explanation as to why disc-tracing CO first overtone (or `bandhead') emission is not a ubiquitous feature in low- to medium-resolution spectra of massive young stellar objects, but instead is only detected toward approximately 25 per cent of their spectra.  In this paper, we investigate the hypothesis that only certain mass accretion rates result in detectable bandhead emission in the near infrared spectra of MYSOs.  Using an analytic disc model combined with an LTE model of the CO emission, we find that high accretion rates ($\gtrsim10^{-4}\,{\rm M}_{\odot}{\mathrm{yr}}^{-1}$) result in large dust sublimation radii, a larger contribution to the $K$-band continuum from hot dust at the dust sublimation radius, and therefore correspondingly lower CO emission with respect to the continuum.  On the other hand, low accretion rates ($\lesssim10^{-6}\,{\rm M}_{\odot}{\mathrm{yr}}^{-1}$) result in smaller dust sublimation radii, a correspondingly smaller emitting area of CO, and thus also lower CO emission with respect to the continuum.  In general, moderate accretion rates produce the most prominent, and therefore detectable, CO first overtone emission.  We compare our findings to a recent near-infrared spectroscopic survey of MYSOs, finding results consistent with our hypothesis.  We conclude that the detection rate of CO bandhead emission in the spectra of MYSOs could be the result of MYSOs exhibiting a range of mass accretion rates, perhaps due to the variable accretion suggested by recent multi-epoch observations of these objects. 
\end{abstract}


\section{Introduction}
\label{sec:intro}

Massive stars play a central role in many areas of astrophysics. As a result, their formation has been the subject of intensive study. In the past, it was not clear whether massive stars could form via disc accretion like their lower-mass counterparts \citep{zinnecker_2007}. Recent radiation and hydrodynamic simulations have demonstrated that in fact, discs are a crucial ingredient of massive star formation \citep{krumholz_2009,kuiper_2011,klassen_2016,rosen_2016,harries_2017}. However, studying these discs is difficult due to the embedded nature of Massive Young Stellar Objects (MYSOs) and the fact that they are generally located at kilo-parsec distances.

\smallskip

There are many indirect indications of discs around MYSOs \citep[see the review of][]{beltran_2016}. Evidence of outflows driven by MYSOs, both in terms of ionized and molecular jets \citep{davis_2004,guzman_2010,purser_2016} and molecular outflows \citep{beuther_2002,zhang_2005,cunningham_2016}, suggest that MYSOs are accreting material. On scales of 1000s of au and larger, there are suggestions of rotating structures that could be feeding a circumstellar disc \citep[so called toroids, e.g.][]{cesaroni_2005, beuther_2008, beltran_2011, maud_2017}, but on smaller scales there are only a handful of objects that appear consistent with discs in Keplerian rotation \citep[e.g.][]{johnston_2015,ilee_2016}.  However, examining the accretion processes that occur on au-scales toward massive young stars using spatially resolved observations is simply not possible with current instrumentation.  Hence, indirect observational methods must be employed.    

\smallskip

Using integral-field spectroscopy, \citet{davies_2010} found evidence that the MYSO W33A drives a bipolar wind traced by Br$\gamma$ emission, and demonstrated the motion of the cool CO material in the vicinity of the star is consistent with Keplerian rotation. In addition, optical and NIR interferometry has provided evidence of flattened structures surrounding MYSOs that are thought to be circumstellar discs \citep{follert_2010,kraus_2010,grellmann_2011,dewit_2011,boley_2013}.

\smallskip

Despite the accumulating evidence for the presence of discs around MYSOs, the properties of such discs, such as the accretion rate through them, remain largely unknown. This is in part because the interferometric observations mentioned above generally trace the distribution of dust in the circumstellar environments of MYSOs, while accretion will occur through an inner gaseous disc. There are few direct tracers of such discs. MYSOs often exhibit Br$\gamma$ emission, a tracer of ionized gas, but it has been shown that this can originate in outflows \citep{davies_2010}, and thus does not provide a unique diagnostic of inner disc conditions. Another emission feature that is seen in the spectra of MYSOs is the ro-vibrational CO $v=2$--0 first overtone (or `bandhead') emission. This emission requires warm ($T =$ 2500--5000\,K) and dense (n $>$ $10^{11}$\,cm$^{-3}$) gas. These are the conditions expected in the inner regions of circumstellar discs and thus this emission has often been used as a tracer of discs \citep{carr_1989,calvet_1991,chandler_1995,najita_1996}.

\smallskip

It has been shown that the appearance of the CO bandhead emission of MYSOs can be reproduced using a model of emission from a circumstellar disc in Keplerian rotation \citep{bik_2004,blum_2004,wheelwright_2010,ilee_2013}. Such studies provided the first evidence of gaseous circumstellar discs surrounding MYSOs. Furthermore, this emission allows us to probe the conditions of such discs for the first time.  In \citet{ilee_2013}, we demonstrate that the properties of the disc models that best fit the CO bandhead emission of MYSOs are generally those of discs of several au in size, interior to the dust sublimation radius and with temperature and surface density gradients approximately consistent with those expected of flat accretion discs. However, we found that simple models of accretion discs were unable to fit the bandhead profiles satisfactorily, instead requiring the freedom of an analytic model to adequately reproduce the emission.  Furthermore, there is currently no explanation as to why CO bandhead emission is not a ubiquitous feature in the spectra of MYSOs. Instead, it is only present in the spectra of approximately 25 per cent of MYSOs \citep{cooper_2013, pomohaci_2017}.  If CO bandhead emission traces discs, it could be inferred that the MYSOs that do not exhibit this emission do not possess discs, contrary to expectations.

\smallskip

It is conceivable that, in order for CO emission to be observed, the inclination of MYSOs is required to be close to face on. However, the distribution of inclinations of MYSOs with CO bandhead emission reported in \citet{ilee_2013} exhibits no preference for low values.  In addition, CO emission has been identified from objects that are known to be viewed close to edge-on (e.g. MWC 349, \citealt{kraus_2000}).  It is also possible that the objects with CO emission represent a different evolutionary stage of MYSOs than those without.  However, \citet{ilee_2013} determined that the objects with and without CO emission appear no different in terms of their NIR and MIR colours, suggesting the objects with CO bandhead emission are representative of the MYSO population as a whole.  Therefore, as yet, the properties of the MYSOs that possess CO bandhead emission do not indicate what dictates whether or not a particular MYSO will exhibit this emission in its spectrum. 

\smallskip

Large scale surveys of massive young stellar objects have shown that their properties such as bolometric luminosity and surrounding extinction can be varied \citep{cooper_2013}.  This is suggestive that the MYSOs that comprise these surveys exhibit a range of stellar properties, and importantly, mass accretion rates.  It has been predicted that models of circumstellar discs of MYSOs can be unstable \citep{krumholz_2007b} and that the accretion rate in these discs can be variable \citep{kuiper_2011}.  Such an interpretation is supported by recent observations of variability in massive young stellar objects, observed at infrared \citep{caratti_2017} and also submillimetre wavelengths \citep{hunter_2017}.  It is therefore reasonable to assume that any given MYSO can exhibit a range of accretion rates across its lifetime.  

\smallskip

In this paper, we assess the hypothesis that only certain accretion rates produce detectable CO bandhead emission, which could then explain the detection rate of CO bandhead emission in the spectra of MYSOs.  We use an analytic model of an accretion disc coupled with our model of CO ro-vibrational emission to examine the effect of changing the accretion rate on CO bandhead emission from such a disc.   Section \ref{sect:modelling} describes the disc model we utilise in this work.  Section \ref{sect:results} details the application of this model to a situation representing an accreting MYSO.  We discuss our findings in Section \ref{sect:discussion} and conclude the paper in Section \ref{sect:conclusions}.

\section{Modelling Approach}
\label{sect:modelling}

\begin{figure}
    \centering
    \includegraphics[width=\columnwidth, trim={0 1cm 0 0},clip]{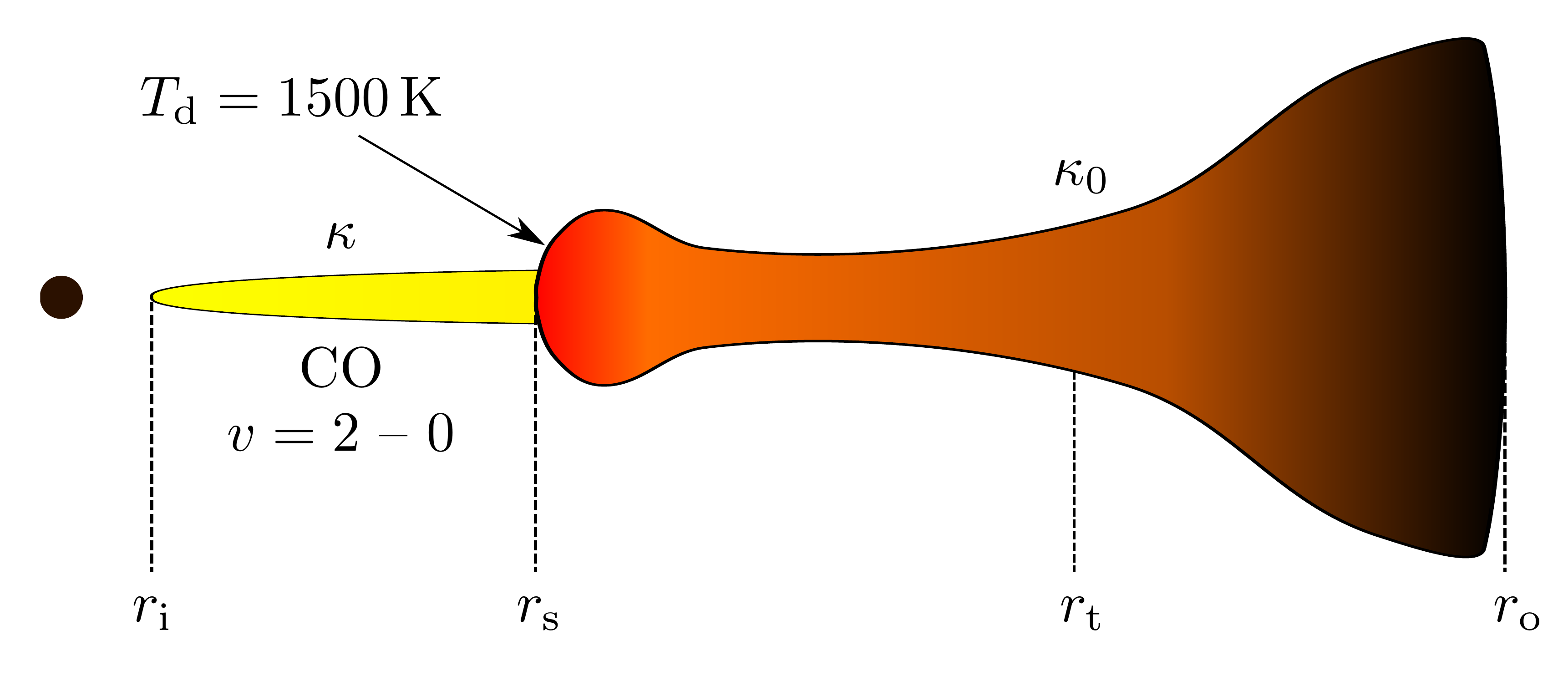}
    \caption{A representative schematic diagram of the disc model considered for this work.  The disc is split into three radial zones.  The first radial zone (between the inner radius $r_{\rm i}$ and the dust sublimation radius $r_{\rm s}$) is considered to be purely gaseous with opacity $\kappa$ (see Equation \ref{eqn:opac}).  The second radial zone (between  $r_{\rm s}$ and the transition radius $r_{\rm t}$) is dominated by viscous heating with opacity $\kappa_{0}$.  The final radial zone (between $r_{\rm t}$ and the outer radius $r_{\rm o}$) is dominated by radiative heating, also with opacity $\kappa_{0}$.  Hot dust close to the sublimation radius is represented by a blackbody emitting at 1500\,K.}
    \label{fig:disc}
\end{figure}

\begin{table}
\centering
    \begin{minipage}{0.8\columnwidth}
    \caption{Star and disc parameters adopted for the model.}
    \label{tab:models}
    \begin{tabular}{lll}
    \hline
Parameter                       &                       &                        \\
    \hline
Stellar mass                    & M$_{\star}$           & 20\,M$_{\odot}$       \\
Stellar radius                  & R$_{\star}$           & 35\,R$_{\odot}$       \\
Stellar luminosity              & L$_{\star}$           & $1.1\times10^{4}$\,L$_{\odot}$\\
Stellar effective temperature   & T$^{\star}_{\rm eff}$         & 10$^{4}$\,K           \\
Distance                        & $d$                   & 2.5\,kpc              \\
Disc outer radius$^{\dag}$      & r$_{\rm o}$           & 200\,au               \\
Disc inner radius               & r$_{\rm i}$           &  0.2\,au                     \\
Disc mass$^{\dag}$          & M$_{\rm disc}$        &   10\,M$_{\odot}$                    \\
Disc inclination                & $i$                   & 0\degr                \\
Disc viscosity                  & $\alpha_{\rm visc}$   & 0.1                   \\
    \hline
    \end{tabular}\\
    \small{$\dag$: Initial value.  These quantities vary as the model evolves. See Appendix \ref{app} for details.}
    \end{minipage}
\end{table}

\begin{figure*}
  \centering
    \includegraphics[width=\textwidth]{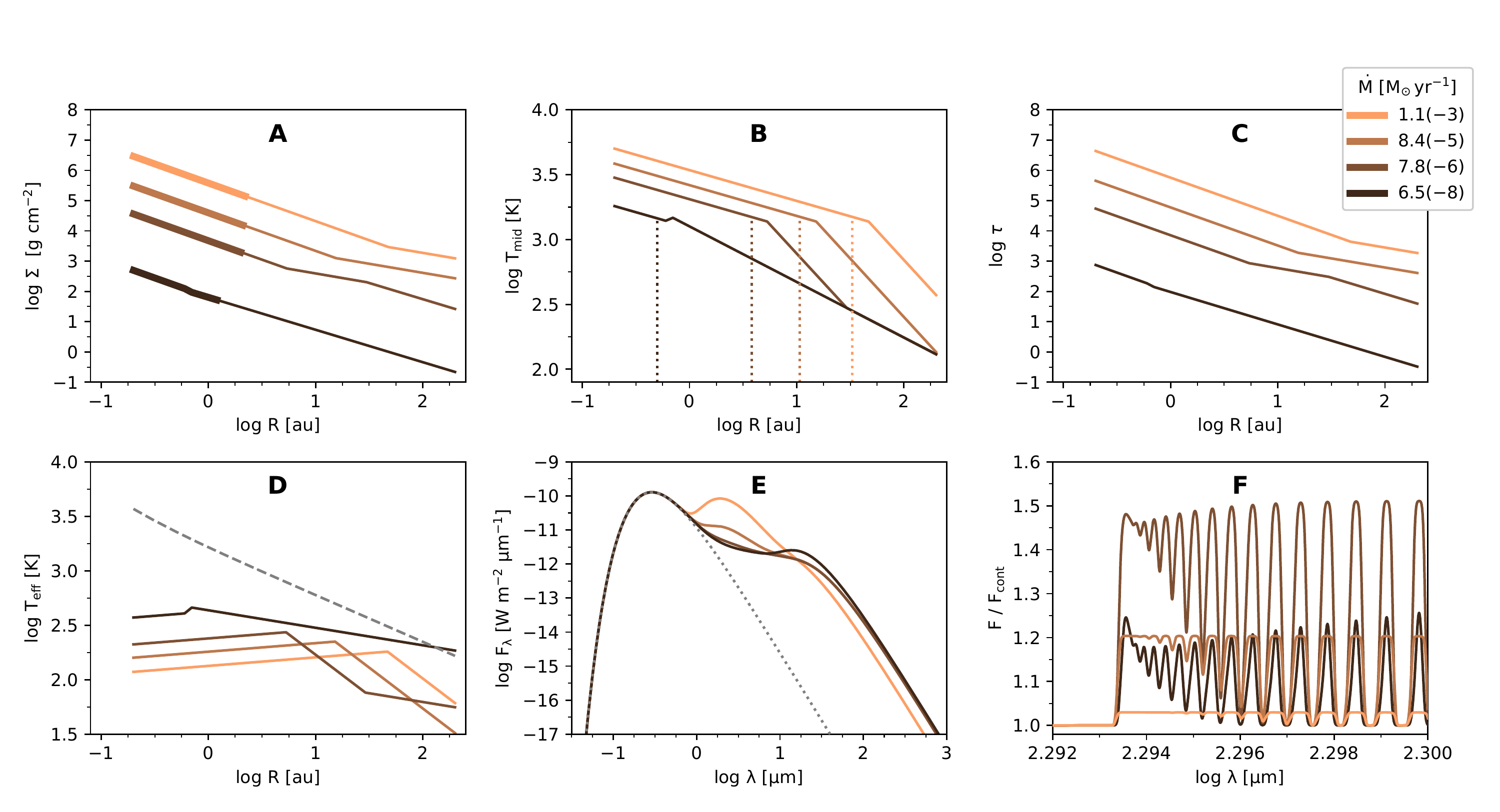}
    \caption{Disc properties and observable signatures for the MYSO model for the range of accretion rates considered. \emph{A)} Disc surface density as a function of radius. \emph{B)} Disc mid-plane temperature as a function of radius, vertical dashed lines indicate the location of the corresponding dust sublimation radius. \emph{C)} Disc optical depth as a function of radius.  \emph{D)} Disc effective temperature, the dashed line indicates the effective temperature profile of a flared, irradiated disc.  \emph{E)} Spectral energy distributions (SEDs), the dashed line indicates the contribution from the central star. \emph{F)} Predicted CO $v=2$--0 first overtone bandhead emission, normalised to the adjacent continuum.  The thicker lines in panel A denote the spatial extent of the region which emits 80 per cent of the CO $v=2$--0 flux for each model.}    
    \label{fig:case_1}
\end{figure*}

Since we aim to assess the continuum and CO bandhead for a range of stellar properties and accretion rates, the ideal model is one that requires minimal computing time. Consequently, we adopt the analytic model reported by \citet{chambers_2009}.   The model describes a disc heated by viscous accretion and irradiation from a central star.  The disc is assumed to be flared and its viscous nature is described by the standard alpha disc model \citep{shakura_1973}.  We use this model to predict disc mid-plane temperatures ($T_{\rm mid}$), effective temperatures ($T_{\rm eff}$) and surface densities ($\Sigma$).  These are then used to predict the resulting continuum emission from the dust in this disc, and also the strength of CO first overtone line emission. 

\smallskip

For a detailed description of the physical disc model, we refer the reader to the comprehensive paper of \citet{chambers_2009}, but here we briefly summarise the key characteristics and modifications made.  The model describes a Keplerian disc with a decreasing mass accretion rate as a function of time. We utilise this to provide different accretion rates for various given model discs, but do not attach significance to the temporal differences between each model.

\smallskip

The disc consists of three main radial zones (see Figure \ref{fig:disc}).  The furthest zone from the central star (between the transition radius $r_{\mathrm{t}}$ and the outer radius $r_{\mathrm{o}}$) is dominated by irradiation from the central star ($Q_{\rm irr} > Q_{\rm visc}$).  The middle zone is located between the dust sublimation radius $r_{\mathrm{s}}$ and $r_{\mathrm{t}}$, and is dominated by viscous heating ($Q_{\rm visc} > Q_{\rm irr}$).  In the former two zones, we adopt a constant opacity of $\kappa_0 = 3.0\, \mathrm{cm^2\,g^{-1}}$ (approximately appropriate for midplane disc temperatures around MYSOs, see \citealt{vaidya_2009}).  The zone closest to the central star is located between the inner disc radius $r_{\mathrm{i}}$ and the dust sublimation radius $r_{\mathrm{s}}$, and describes an inner gaseous disc.  Within this zone, dust is entirely sublimated and the opacity is assumed be dependent on temperature, following 
\begin{equation}
\kappa = \kappa_0(\frac{T}{T_e})^n,
\label{eqn:opac}
\end{equation}
where $T_e = 1380$~K and $n=-14$, based on the analytic opacity fits of \citet{stepinski_1998}.  Throughout all zones in the disc, we assume an adiabatic index and mean molecular weight to $\gamma = 1.4$ and $\mu = 2.4$ respectively. 

\smallskip

To calculate the CO bandhead emission from such a disc, we utilise our model of CO emission \citep[previously applied in][]{wheelwright_2010, ilee_2013, murakawa_2013, ilee_2014}.  The model considers local thermodynamic equilibrium, and requires the excitation temperature and number density of the CO at the location of emission.  The surface density of CO molecules is calculated from the disc mass surface density assuming a fractional abundance of $1\times10^{-4}$ for CO. In principle, the effective temperature of the disc is calculated using
\begin{equation}
T_{\rm{eff}}^4 = \frac{4}{3 \tau}T_{\mathrm{mid}}^4.
\end{equation}
In the cases of high accretion we consider here, the optical depth $\tau$ is very large. Consequently, the effective temperature due to the mid-plane temperature is low. In these cases, where the effective temperature is lower than the temperature due to stellar irradiation, we adopt the temperature due to stellar irradiation as the excitation temperature of the CO molecules at the surface of the disc \citep[following][]{chiang_1997}. Since the CO emission is thus placed at the surface of the disc, we assume that it is optically thin, i.e. if the surface density is sufficient to produce optically thick emission we set it to a value that produces only marginally thick emission ($N_{\rm CO} = 5\times 10^{21}~\mathrm{cm^{-2}}$).

\smallskip

We split the disc into 500 rings in the radial direction and the CO emission of each ring is calculated based on the local temperature as described above.  For each rotational transition of CO, we assume a Gaussian linewidth of 15\,km\,s$^{-1}$ (a typical value obtained from model fits to high resolution spectra of CO overtone emission toward massive YSOs and Herbig Be stars, see \citealt{ilee_2013,ilee_2014}).  As in our previous work, we only include emission from regions between 1000 -- 5000\,K, as higher temperatures will result in the dissociation of the CO molecule, and lower temperatures will produce negligible levels of emission. These spectra are then added cumulatively in order to predict the total spectrum for the entire disc of CO.  

\smallskip

To estimate the underlying continuum, we assume that each disc ring emits as a black-body at the same temperature we estimate for the CO. This assumes that the disc surface appears continuous. However, the appearance of discs around MYSOs is likely strongly influenced by dust close to the dust sublimation radius \citep[see e.g.][]{kraus_2010}. To approximate the emission of dust at the dust sublimation radius, we also include a ring of emission located at the dust sublimation radius (where the mid-plane temperature is 1500~K) that has a width of 10 per cent of its radius, and emits as a black body at 1500~K. This is of course an over simplification, and the true sources of continuum emission in MYSOs will involve many other aspects.  Nevertheless, our approach offers a first step in the exploration of how the ratio of CO and continuum emission may change as a function of accretion rate in these objects.

\subsection{Star-disc properties}

It is expected that MYSOs that are heavily accreting will initially be relatively cool before emerging on the ZAMS as hot stars \citep[see, e.g.][]{hosokawa_2009}.  In fact, \citet{pomohaci_2017} were recently able to spectral-type the MYSO G015.1288$-$0.6717, finding that the star is cooler and larger than what might have been expected based on its total luminosity alone.  Hence, our chosen parameters that describe the central star and disc reflect this, corresponding to an MYSO that is relatively distant from the ZAMS with a large stellar radius (35\,R$_{\odot}$) and low effective temperature ($10^{4}$\,K) due to rapid accretion.  The stellar mass was set at 20\,M$_{\odot}$, and the luminosity was thus $1.1\times10^{4}$L$_{\odot}$.  The initial disc mass is set to half of the stellar mass (10\,M$_{\odot}$), which decreases quickly in our model due to accretion (see Appendix \ref{app}).  The initial disc radius was set to 200\,au, but the disc spreads viscously outwards with time.  Larger discs simply affect the millimetre portion of the spectral energy distribution, and not the near-infrared portion, which is important for our studies.  The viscosity parameter within the disc was set to $\alpha_{\rm visc} = 0.1$. Arguably, the value of $\alpha$ is not critical as the accretion history of MYSOs is not strongly dependent on this parameter \citep{kuiper_2011}.  In order to simplify the number of free parameters involved in the modelling, and the fact that we do not here attempt to model any given system, we assume that we view the disc in a face-on orientation ($i=0\degr$) and at a distance of 2.5\,kpc in all models presented.  We summarise our chosen star-disc properties in Table \ref{tab:models}.

\section{Results}
\label{sect:results} 

Figure \ref{fig:case_1} depicts the disc properties and observables that have been calculated for the MYSO model described in Section \ref{sect:modelling}.  For each panel, four accretion rates are shown -- $1.1 \times 10^{-3}$, $8.4 \times 10^{-5}$, $7.8 \times 10^{-6}$ and $6.5 \times 10^{-8}$\,${\rm M}_{\odot}{\mathrm{yr^{-1}}}$.  Panel A shows the disc surface density as a function of radius.  Panel B shows the disc mid-plane temperature as a function of radius, within which vertical dashed lines correspond to the radial location of the dust sublimation radius (assumed to be 1500\,K).  Panel C shows the disc optical depth $\tau = \frac{\kappa \Sigma}{2}$ as a function of radius.  Panel D gives the effective temperature of the disc as a function of radius, in which the dashed line indicates the effective temperature of a flared, irradiated disc.  Panel E depicts the spectral energy distributions (SEDs) for each disc model considered, where the contribution from the central star is shown with a dashed line.  Panel F shows the resulting CO $v=2$--0 first overtone emission for each disc model, which has been normalised to the adjacent $K$-band continuum.

\smallskip

For all accretion rates considered, high surface densities (panel A) create high optical depths (panel C) that result in relatively low disc effective temperatures ($\la 400$~K, panel D).  Consequently, the disc surface temperature is dominated by stellar irradiation within the inner 100~au (dashed grey line).  High accretion rates result in high midplane temperatures and large dust sublimation radii (panel B).  Consequently, the ring representing dust emission at the sublimation radius is relatively large, contributing more and more flux to the SED (panel E).   In combination, these effects lead to changes in the ratio of the CO first overtone flux to the continuum flux as a function of mass accretion rate, from 1.05 to 1.5 times the level of the adjacent continuum (panel F).  

\smallskip

\begin{figure}
  \centering  
    \begin{minipage}{0.95\columnwidth}
    \includegraphics[width=\columnwidth]{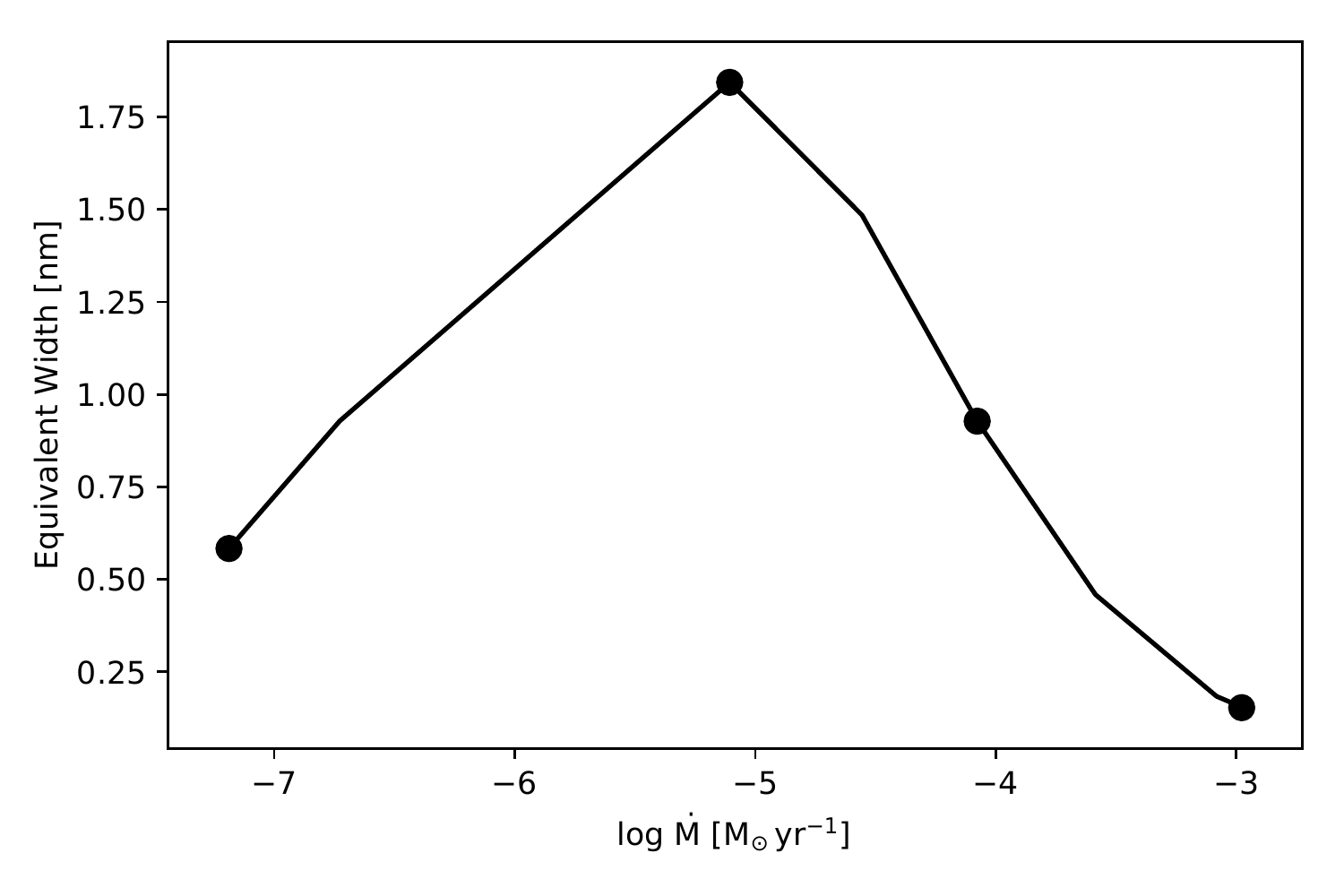}
    \caption{Absolute value of the equivalent width of the CO v=2=0 transition measured for each of model spectra as a function of mass accretion rate, with filled points representing the mass accretion rates shown in Figure \ref{fig:case_1}.  The strongest lines are seen for intermediate accretion rates.  Equivalent widths were measured between 2.29 and 2.30\,\micron\ from the continuum normalised spectra.} 
    \label{fig:equivalent_widths} 
    \end{minipage}
\end{figure}

The highest accretion rates, $1.1\times10^{-3}$~$M_{\odot}{\mathrm{yr^{-1}}}$, are associated with relatively low line-to-continuum-ratios ($\la$1.05) due to the large amount of flux from the ring of dust emission, which is much larger in size than the region from which the CO emission originates.  In addition, due to the high temperatures and high gas densities in these models, a larger fraction of rotational transitions close to the bandhead peak are excited and optically thick, leading to a flatter overall bandhead profile.   Conversely, but displaying a similar overall result, low accretion rates, $6.5\times 10^{-8}$~$M_{\odot}{\mathrm{yr^{-1}}}$, are also associated with lower CO line-to-continuum ratios ($\la$1.2) due to the drop in surface density of approximately 2 orders of magnitude and the smaller size of the CO emitting region. As the mass accretion rate decreases, rotational transitions close to the bandhead peak become less optically thick, reducing the flattening effect and recovering the typical bandhead shape.  Somewhat counter-intuitively, the strongest line-to-continuum ratios ($\sim 1.5$) are associated with intermediate accretion rates, due to the interplay between the strength of the continuum emission mentioned above.  Figure \ref{fig:equivalent_widths} shows the resulting measured equivalent widths of the CO $v=2$--0 emission line for our series of models as a function of mass accretion rate, clearly showing that the largest equivalent widths are associated with intermediate accretion rates, of order $10^{-5}$\,$M_{\odot}{\mathrm{yr^{-1}}}$.

\subsection{Simulated observations}

We now investigate how the emission from our series of models would be observed given typical spectroscopic observing campaigns of massive young stellar objects.   A detection rate for CO first overtone emission of 17 per cent was determined from the near-infrared spectroscopic survey of 247 objects of \citet{cooper_2013} drawn from the Red MSX Source (RMS) survey\footnote{\url{http://rms.leeds.ac.uk}} \citep{lumsden_2013}.  Typically, these data possessed a signal-to-noise-ratio of between 50--100 with a spectral resolution of $R = \Delta \lambda / \lambda \sim 500$.  Further studies of the near-infrared spectra of MYSOs have employed spectrometers with a higher spectral resolving power, but these have been limited to a smaller sample size and may be targeted toward objects with confirmed CO emission, rather than being unbiased surveys.  Medium resolution spectra ($R\sim7000$) of 36 MYSOs drawn from the RMS survey were presented by \citet{pomohaci_2017}, with a typical signal-to-noise ratio of 100, leading to a detection rate of CO first overtone emission of 34 per cent.

\smallskip 

In order to assess the observability of the models presented in Figure \ref{fig:case_1}, we utilise several tasks within the Image Reduction and Analysis Facility (IRAF\footnote{\url{http://iraf.noao.edu}}), which we outline here.  Firstly, the model spectra are convolved with a Gaussian profile whose full-width at half maximum (FWHM) corresponds to the resolution element that produces the desired spectral resolution $R = \Delta \lambda / \lambda$.  Next, the convolved spectra are re-binned such that the number of points,  $N_{\rm pts}$, across the $v=2$--0 bandhead is comparable to that of the observational data.  Finally, uniform random noise is added to the spectrum in order to reproduce the approximate signal-to-noise ratio required.  Table \ref{tab:symobs} shows the parameters we have adopted for each case of simulated observations we consider.

\smallskip

\begin{figure*}
  \centering  
    \includegraphics[width=\textwidth]{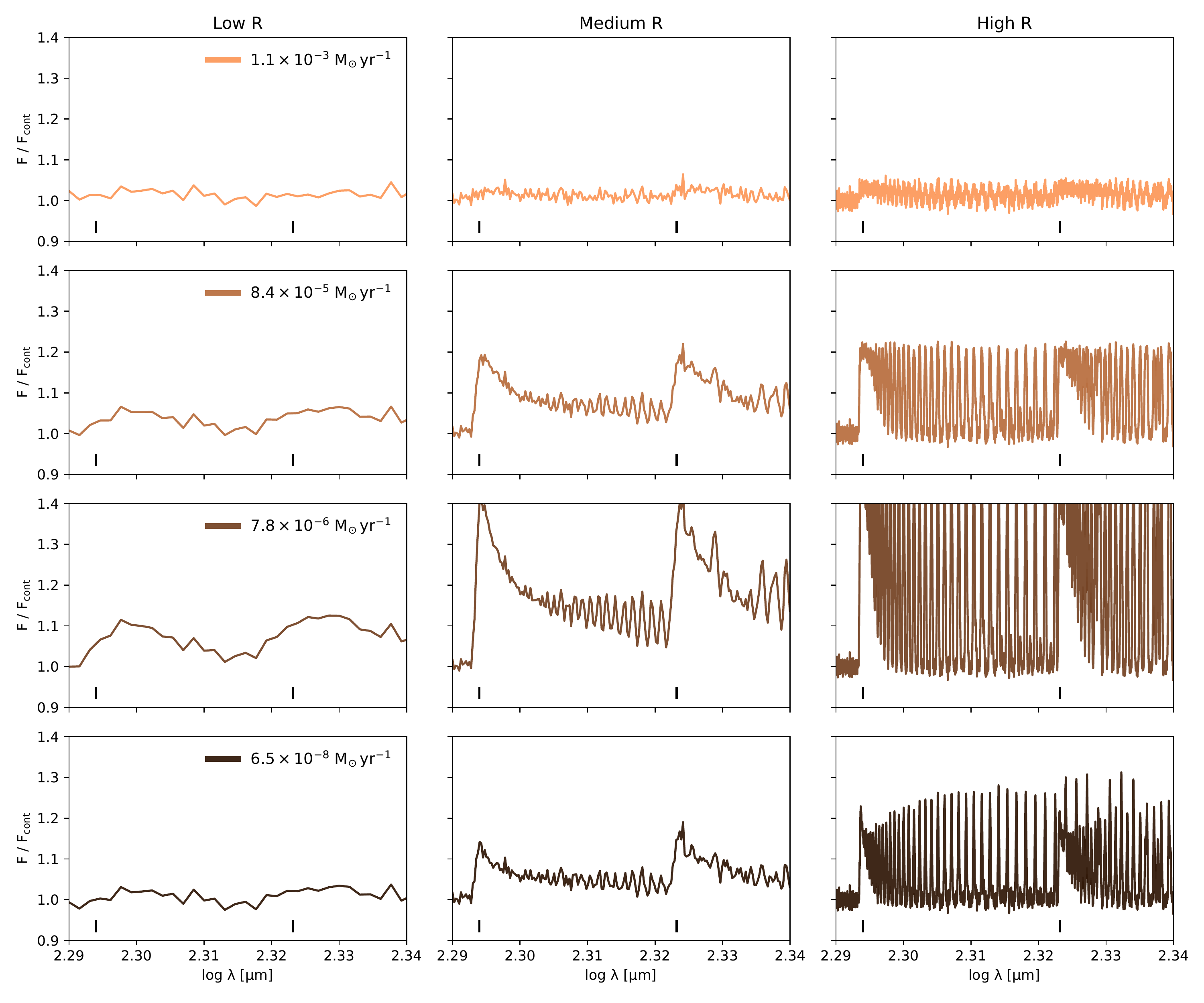}
    \caption{Simulated observations of the CO spectra presented in Figure  \ref{fig:case_1}, with a spectral resolution $R = 500$, 7000 and 50,000 (left, middle, right respectively).  All spectra possess the same signal-to-noise ratio of approximately 100.  Vertical ticks mark the wavelengths of the $v=2$--0 and 3--1 transitions in each panel.  Only moderate accretion rates result in detectable emission for all spectral resolutions considered, and very high spectral resolution is required in order to ensure detection for all accretion rates considered.}
    \label{fig:case_1_co_spec} 
\end{figure*}

Figure \ref{fig:case_1_co_spec} depicts the resulting simulated spectra for these models  along with vertical tick marks to show the wavelengths at which the $v=2$--0 and 3--1 bandheads should be visible.  As can be seen, for the spectra corresponding to both the highest and lowest accretion rates ($1.1 \times 10^{-3}$ and $6.5 \times 10^{-8}$\,${\rm M}_{\odot}{\mathrm{yr^{-1}}}$, respectively) show no detectable signs of CO emission in low resolution spectra (left).  However, models with  intermediate accretion rates ($8.4 \times 10^{-5}$ and $7.8 \times 10^{-6}$\,${\rm M}_{\odot}{\mathrm{yr^{-1}}}$) result in CO emission that is detectable in spectra of this signal-to-noise and spectral resolution.  At medium spectral resolution (middle), the highest accretion rates are still undetectable, but lower accretions rates, $\lesssim 10^{-6}$\,${\rm M}_{\odot}{\mathrm{yr^{-1}}}$, result in detectable emission.  Finally, we examine the effect of observing our models with an extremely high resolution spectrograph resulting in a resolving power of $R \sim 50\,000$, shown in Figure \ref{fig:case_1_co_spec} (right).  With data of these quality, it is possible to detect the CO $v=2$--0 and $v=3$--1 bandheads for all mass accretion rates considered in our models.  Based on these results, it is clear that a combination of both high spectral resolution and high signal to noise would be required to detect all levels of CO first overtone emission in massive young stellar objects undergoing accretion in the manner described by our modelling.

\begin{table}
    \centering
    \begin{minipage}{0.65\columnwidth}
    \caption{Parameters adopted for each of the \hspace{\textwidth} simulated observation cases.}
    \label{tab:symobs}
    \begin{tabular}{lccc}
    \hline
Case                &     Low $R$    &   Medium $R$  &   High $R$\\
    \hline
    $R$             &   500          & 7000      & 50,000       \\            
    $N_{\rm pts}$   &   20           & 120       & 2000         \\
    SNR             &   100          & 100       & 100          \\
    \hline
    \end{tabular}
    \end{minipage}
\end{table}

\section{Discussion}
\label{sect:discussion}

Using an analytic model of a viscous alpha disc that is also irradiated by a central star, we have shown that discs with high accretion rates result in large optical depths which result in relatively low disc effective temperatures. In most of the cases considered, the effective temperature due to stellar irradiation is greater than that predicted by the viscous model. This highlights that it is difficult to constrain accretion rates using CO bandhead emission if the accretion rates in question are high.  Such a result concurs with the findings of \citet{ilee_2013} in which simple models of accretion discs \citep[e.g.][]{carr_1989, chandler_1995}  were unable to satisfactorily reproduce high resolution spectra CO first overtone emission toward MYSOs.

\smallskip

Our predictions for the strength of the CO bandhead emission in low resolution spectra do not exceed 20 per cent of the underlying continuum.  This is in agreement with the near-infrared spectroscopic survey of MYSOs presented in \citet{cooper_2013}, in which detections range from 5 to 20 percent of the adjacent continuum for a spectral resolving power of $R \sim 500$.  Another consequence of the disc surface temperature being dominated by stellar irradiation is that the surface temperature of discs around MYSOs is unlikely to be less than that of the adjacent interior, meaning that it is unlikely CO bandhead appears in absorption.  Such a finding is also in agreement with \citet{cooper_2013}, in which only two objects exhibit CO first overtone absorption.  In addition, these objects (G023.6566$-$00.1273 and G032.0518$-$00.0902) exhibit relatively low luminosities of $\sim 5\times10^{3}$\,L$_{\odot}$, and as such may be FU Ori-type objects undergoing an outburst, rather than bona fide MYSOs.   

\smallskip

It is not clear why CO first overtone emission is not a ubiquitous feature in the spectra of MYSOs.  From our findings here, we hypothesise that since the strength of CO bandhead emission in our model is dependent on accretion rate, the detection rate of CO bandhead emission in the spectra of MYSOs reflects the fact that MYSOs exhibit different accretion rates.  We show that different accretion rates can produce a range of CO bandhead to continuum ratios ranging from approximately 1.05 to 1.5. If the accretion rates of MYSOs early in their evolution is variable between high and moderate values ($\sim 10^{-4}-10^{-5}\,{\rm M}_{\odot}{\mathrm{yr}}^{-1}$), this scenario can account for some of these objects exhibiting CO bandhead emission and others appearing not to (when observed with a low spectral resolution).  A full analysis of this scenario will require detailed radiative transfer calculations including both the gas and the dust in the environments of MYSOs \citep[see e.g.][]{ercolano_2013}. However, our simplistic model demonstrates that this scenario is at least plausible.

\smallskip

We note that, in this case of high accretion rates, the CO bandhead emission is largely overwhelmed by the continuum emission originating in the ring at the dust sublimation radius. This means that if this scenario is correct, the objects that do not exhibit CO bandhead emission due to rapid accretion should possess a significant continuum excess due to hot dust.   For our models, the $K$ band flux that originates in the ring at the dust sublimation radius is approximately 85 per cent of the total flux for a mass accretion rate of $1.1\times 10^{-3}\,{\rm M}_{\odot}{\mathrm{yr}}^{-1}$ and falls to approximately 2 per cent of the total when the mass accretion rate is $7.8 \times 10^{-7}\,{\rm M}_{\odot}{\mathrm{yr}}^{-1}$. Consequently, assessing the contribution of hot dust to the continuum flux of MYSOs could allow us to test our hypothesis. 

\smallskip

The NIR colours of the MYSOs that exhibit CO bandhead emission appear no different to those that do not \citep{ilee_2013}. This appears contrary to the prediction that CO emission is associated with objects that do not have a significant dust excess. However, \citet{ilee_2013} do not account for different reddening, which is likely to be dependent on the circumstellar environment. Furthermore, the predicted difference between high and moderate accretion states in terms of the $J-K$ colour is not very large (approximately 1 magnitude). Therefore, that the NIR colours of MYSOs with and without CO bandhead emission appear similar does not yet exclude our hypothesis. A detailed study of the reddening towards a large sample of MYSOs is required to further examine this possibility.  However, we note that \citet{porter_1998} examined the reddening and dust contribution in the NIR spectra of 12 luminous young stellar objects. Interestingly, they found that the object with the smallest dust excess, IRAS 17441$-$2910, exhibited the strongest CO bandhead emission, which is in agreement with the scenario outlined above. 

\subsection{Comparison to a recent spectroscopic survey of MYSOs}
\label{sec:obs}

We draw upon the recent study of \citet{pomohaci_2017} in order to test whether our hypothesis that only intermediate mass accretion rates result in detectable CO first overtone emission in massive young stellar objects.  Spectra were obtained from 1.07 to 2.33\,\micron\ using GEMINI-North/GNIRS with a spectral resolving power of $R\sim7000$.  Such a wavelength coverage allowed simultaneous observation of the CO first overtone bandheads ($v=2$--0 and 3--1) alongside the Br\,$\gamma$ hydrogen recombination emission line.  The luminosity of many emission lines, including Br\,$\gamma$, has been shown to correlate with the accretion luminosity in many low and intermediate mass young stellar objects  \citep[e.g.][]{mendigutia_2011,mendigutia_2015}, and as such, can be used to calculate the accretion rate onto the central object. 

\smallskip

\begin{figure}
  \centering  
    \begin{minipage}{0.95\columnwidth}

    \includegraphics[width=\columnwidth]{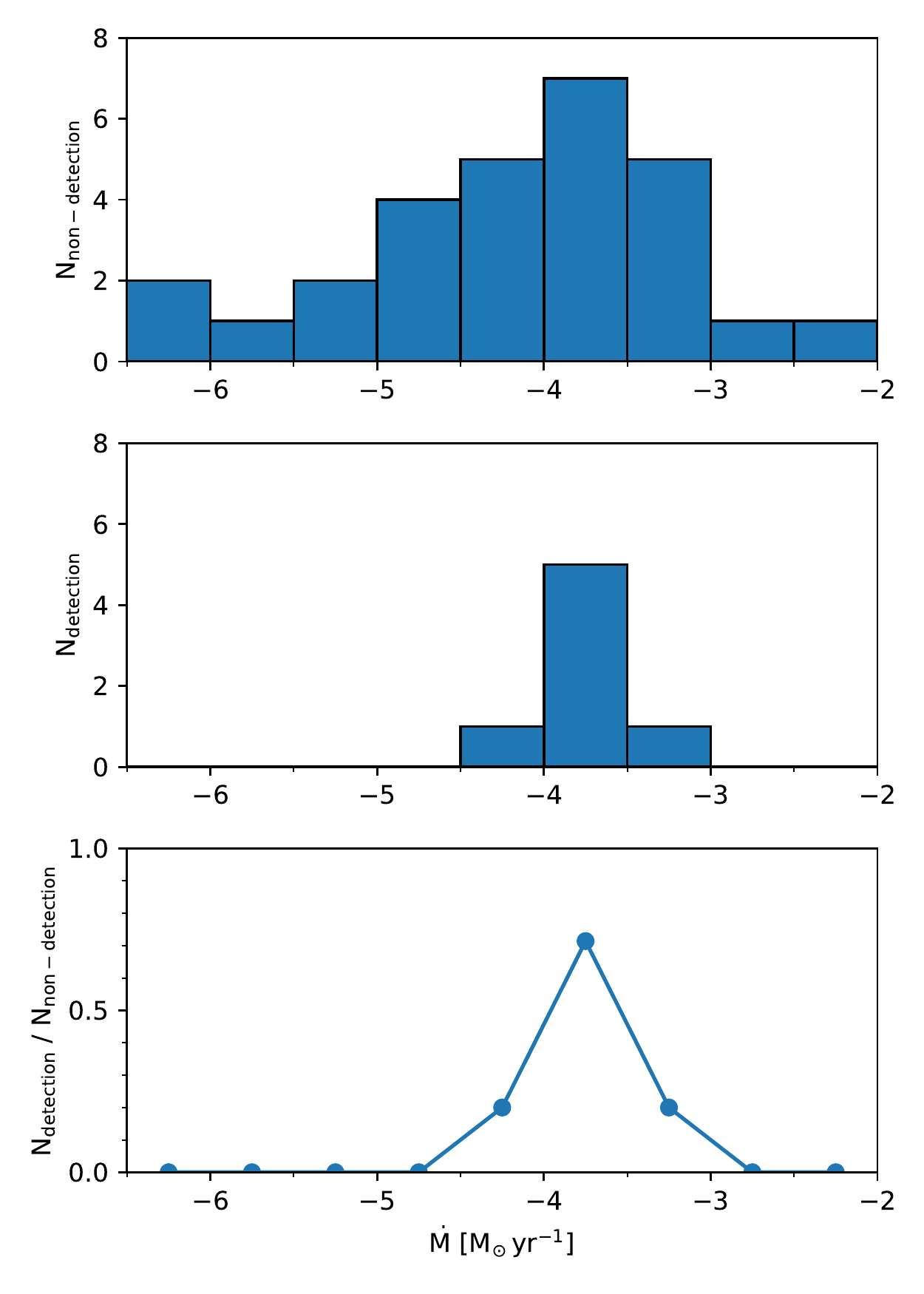}
    \caption{Mass accretion rates calculated from the luminosity of the Br\,$\gamma$ line for the 38 MYSOs with $R \sim 7000$ presented in \citet{pomohaci_2017} .  The upper panel shows objects where CO first overtone emission is not detected in their spectra, the middle panel shows objects where CO first overtone emission is detected, and the relative fraction of detections against non detections for each mass accretion rate bin are shown in the lower panel.  The CO detections are confined to a narrower range of accretion rates than the non-detections, and the relative fraction of detections compared with non-detections peaks for intermediate values of accretion rate ($\sim 2\times10^{-4}$\,M$_{\odot}{\mathrm{yr}}^{-1}$).} 
    \label{fig:gemini_co_mdots} 
    \end{minipage}
\end{figure}

Figure \ref{fig:gemini_co_mdots} shows the accretion rates calculated from the Br\,$\gamma$ line luminosities for the 38 MYSOs presented in \citet{pomohaci_2017}.  The upper panel shows those in which no CO first overtone emission was detected, the middle panel shows those in which CO first overtone emission was detected, and the lower panel shows the relative fraction of detections to non-detections for each mass accretion rate bin.  The CO first overtone non-detections span a wide range of accretion rates, $1.4 \times 10^{-2}$ -- $6.9 \times 10^{-7}$\,M$_{\odot}{\mathrm{yr}}^{-1}$. The CO first overtone detections, while less numerous, span a much smaller range of intermediate mass accretion rates $5.5\times10^{-4}$ -- $5.1\times10^{-5}$\,M$_{\odot}{\mathrm{yr}}^{-1}$, which is consistent with our hypothesis.  While both distributions display a peak at intermediate accretion rates, it is interesting to note that the relative fraction of detections to non-detections increases significantly for intermediate values of accretion rate at this peak (centred on approximately $2\times10^{-4}$\,M$_{\odot}{\mathrm{yr}}^{-1}$).  
\smallskip

We note that the errors associated in the determination of accretion rates from observations of emission lines are large, and therefore these results are likely subject to significant scatter.  In addition, we stress that the modelling presented here is far from exhaustive given the complex nature of the circumstellar environment of massive young stellar objects.  In particular, the precise accretion rates that we derive for each model, and the strength of the underlying dust continuum,  will be influenced by many of the parameters we have chosen to fix during the modelling process.  Nevertheless, our approach offers the first analysis of the interplay between continuum emission and CO overtone emission in massive young stellar objects, and our results are consistent with our hypothesis that only intermediate values of accretion rate result in detectable levels of CO first overtone emission in low-to-medium resolution spectra.  High spectral resolution observations of a large number of MYSOs, involving simultaneous observation of accretion tracing lines such as Br\,$\gamma$ alongside the CO first overtone emission feature, will be essential in further testing our hypothesis and reducing the errors associated with the determination of accretion rates from these observations.

\section{Conclusions}
\label{sect:conclusions}

In this paper we examine possible explanations for the fact that while all MYSOs are expected to possess circumstellar discs, only a subset ($\sim$25 per cent) exhibit CO bandhead emission, which has been shown to trace the presence of small scale gaseous discs around these objects.  In particular, we investigate the hypothesis that the strength of the CO bandhead emission of MYSOs is sensitive to their accretion rate, and that only particular accretion rates result in observable emission.  We summarise our findings as follows:  

\begin{itemize}
    \item High accretion rates ($>10^{-4}{\rm M}_{\odot}{\mathrm{yr}}^{-1}$) are associated with high surface densities which in turn prevent the CO at the disc surface being heated by the accretion. In addition, high accretion rates result in optically-thick CO bandheads, large dust sublimation radii, and continuum emission that is stronger than the bandhead emission. On the other hand, low accretion rates ($<10^{-6}{\rm M}_{\odot}{\mathrm{yr}}^{-1}$) result in low surface densities and also produce low bandhead emission. In general, moderate accretion rates ($\sim 10^{-5}{\rm M}_{\odot}{\mathrm{yr}}^{-1}$) produce the most prominent CO bandhead emission.
    
    \smallskip
    
    \item We demonstrate that both high signal-to-noise ($>100$) and high spectral resolution ($R>50\,000$) near infrared observations are required in order to detect the relatively weak CO first overtone emission that may be masked in objects exhibiting the highest accretion rates. 
    
    \smallskip
    
    \item We compare our results to findings of a recent medium resolution spectroscopic survey of MYSOs.  We show that the proportion of objects with detected CO first overtone emission peaks for intermediate mass accretion rates, which is consistent with our hypothesis. 
    
    \smallskip
    
    \item We conclude that the detection rate of CO bandhead emission in the spectra of MYSOs ($\sim$25 per cent) could be the result of MYSOs exhibiting a range of accretion rates, either due to an intrinsic spread of accretion rates in these objects, or variable levels of accretion for individual objects as a function of time.  

\end{itemize}

\smallskip

Based on our findings, the non-ubiquitous detection of CO first overtone emission in the spectra of MYSOs may suggest that the majority of these objects surveyed are experiencing high accretion rates, which act to mask their CO bandhead emission.  Forthcoming instruments, such as CRIRES$+$ on the VLT \citep{follert_2014}, will enable high spectral resolution and high signal-to-noise observations of CO overtone spectra, which we have shown is essential in order to detect the relatively weak emission from highly accreting objects.  In addition, the large wavelength coverage of CRIRES$+$ will enable a simultaneous observations of the Br\,$\gamma$ emission line, an key diagnostic of accretion rates \citep[see e.g.][]{mendigutia_2011,fairlamb_2015, fairlamb_2017}.  Such observations will be essential in allowing us to further test our hypothesis, and shed light on the accretion processes and immediate circumstellar environment of massive young stars.

\section*{Acknowledgements}

We would like to thank the anonymous referee for a thorough report which greatly improved this manuscript. JDI gratefully acknowledges support from the DISCSIM project, grant agreement 341137, funded by the European Research Council under ERC-2013-ADG.  RP gratefully acknowledges the studentship funded by the Science and Technologies Facilities Council of the United Kingdom.  This paper made use of information from the Red MSX Source survey database at \url{http://rms.leeds. ac.uk/} which was constructed with support from the Science and Technology Facilities Council of the UK, and NASA's  Astrophysics  Data  System  Bibliographic Services.

\bibliographystyle{mn2e} 
\bibliography{bib}

\appendix

\section{Model verification}
\label{app}

\begin{figure}
  \centering
    \includegraphics[width=\columnwidth]{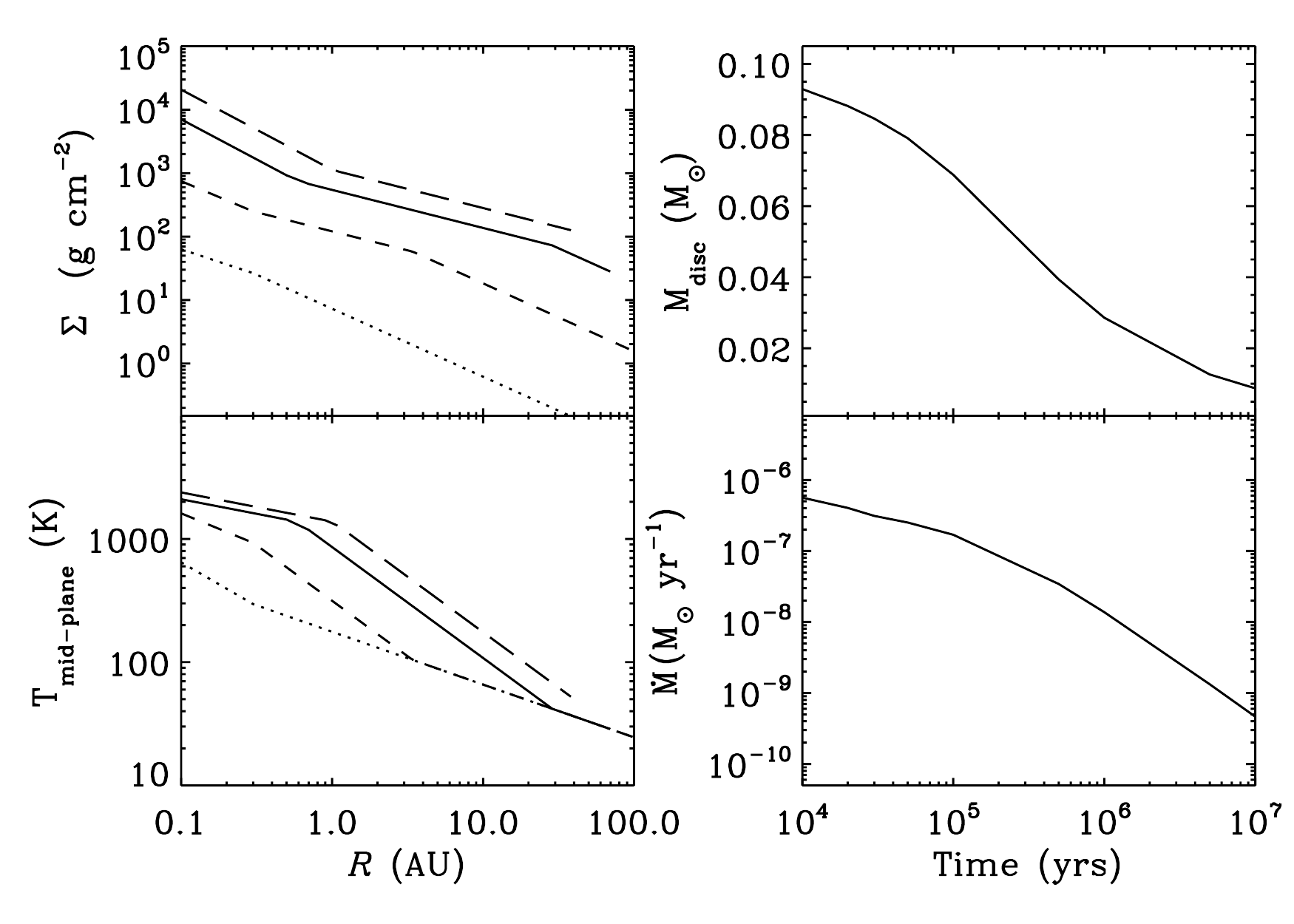}
    \caption{Disc properties from one of the examples presented by \citealt{chambers_2009} (their Figure 1), calculated with our implementation of the model.   The long-dashed line corresponds to a time of $\mathrm{10^4}$~yrs, the solid line to a time of $\mathrm{10^5}$~yrs, the short-dashed line a time of $\mathrm{10^6}$~yrs and the dotted line a time of $\mathrm{10^7}$~yrs.  These show good agreement with the results of \citet{chambers_2009}, demonstrating our implementation of the model is accurate.}
    \label{fig:model_ex}
\end{figure}

To demonstrate our implementation of the model, we consider an example presented by \citet{chambers_2009}. The central star mass is set to $M_{\star}=1~M_{\odot}$ with an effective temperature and radius given by $T_{\mathrm{eff}}=4200$~K and $R_{\star}=3R_{\odot}$.  The initial disc is characterised by an outer radius $s_0 = 33$~AU and a mass of $M_0 = 0.1~M_{\odot}$, respectively. The viscosity is set as $\alpha = 0.01$ and the opacity is given by $\kappa_0=3~{\mathrm{cm^2~g^{-1}}}$.  The disc properties and their subsequent evolution are shown in Fig. \ref{fig:model_ex}.  The surface density decreases with time as material is accreted onto the central star. The various power laws exhibited by the surface density and temperature trace, in increasing distance from the central star, the `evaporative', `viscous' and `irradiated' zones in the disc (where evaporative refers to the sublimation of dust grains, and thus a purely gaseous inner disc).  The viscous region possesses the steepest temperature gradient, and temperatures in all zones decrease with time (or accretion rate), but do so more slowly than surface density. Both disc mass and accretion rates decrease with time. There is a slight break in the slope of accretion rate where the zone dominated by stellar irradiation first appears. After this point, it is assumed the disc evolution is governed by the irradiation. All these points and the plots in Fig. \ref{fig:model_ex} are in good agreement with the findings of \citet{chambers_2009}.

\bsp	
\label{lastpage}
\end{document}